# Impact of hydrogenation on the structure, chemistry, and electrical properties of flame-synthesized carbon nanoparticle films

Luca Basta[a,b,*], Francesca Picca[b], Pegah Darvehi[c], Vincenzo Pagliara[d], Alberto Aloisio[a], Mario Commodo[c,*], Patrizia Minutolo[c], Vito Mennella[d], Stefan Heun[e], Stefano Veronesi[e], and Andrea D'Anna[b]

[a]*Dipartimento di Fisica Ettore Pancini, Università degli Studi di Napoli Federico II, Complesso Universitario di Monte S. Angelo, Edificio 6, Via Cinthia, 21, Napoli 80126, Italy*
[b]*Dipartimento di Ingegneria Chimica, dei Materiali e della Produzione Industriale, Università degli Studi di Napoli Federico II, P.le Tecchio 80, Napoli 80125, Italy*
[c]*Istituto di Scienze e Tecnologie per l'Energia e la Mobilità Sostenibili, Consiglio Nazionale delle Ricerche, P.le Tecchio 80, Napoli 80125, Italy*
[d]*Istituto Nazionale di Astrofisica - Osservatorio Astronomico di Capodimonte, Via Moiariello 16, Napoli 80131, Italy*
[e]*NEST, Istituto Nanoscienze - CNR, and Scuola Normale Superiore, Piazza San Silvestro 12, 56127 Pisa, Italy*

___

**Abstract**

The interaction between hydrogen atoms and carbon nanoparticles is a fundamental process governing the properties of carbonaceous materials in environments ranging from combustion systems to the interstellar medium, and yet not fully understood. This study investigates the effects of controlled atomic hydrogen exposure on young and mature soot nanoparticles, generated in premixed ethylene-air flames, and deposited on substrates. We employed a multi-technique approach to characterize the chemical, mechanical, and electrical evolution of the films. In-situ infrared spectroscopy revealed non-monotonic behavior: an initial increase in aliphatic CH bonds was observed, followed by a decrease at higher hydrogen fluences, a trend more pronounced in the younger, more reactive soot film. This was accompanied by a continuous decrease in the aromatic C=C signal. Atomic force microscopy showed a significant increase in the Young's modulus of the film for both sample types after hydrogenation, indicating increased stiffness. This mechanical change was correlated with an increase in the $I(D)/I(G)$ ratio from Raman spectroscopy, suggesting structural modification. Furthermore, both macroscopic current vs. voltage and local scanning tunneling spectroscopy measurements demonstrated a notable increase in electrical conductivity. For single just-formed soot particles, moreover, a hydrogen-induced transformation from a semiconductive to a semi-metallic nature was observed, with the closing of the electronic band gap. The collective evidence points towards an H-induced CC cross-linking mechanism within the nanoparticle films. We propose that atomic hydrogen facilitates the formation of radical sites, which promotes covalent bond formation between adjacent particles or molecular units, creating a more interconnected and rigid network, with smaller interlayer distance. These findings provide crucial insights into the structural evolution of carbonaceous materials in hydrogen-rich environments, with direct implications for understanding soot formation and for the tailored design of carbon-based materials.

___

*Corresponding authors.

# 1. Introduction

Carbon-based nanomaterials, distinguished by their exceptional electronic, optical, and mechanical properties, have been the subject of intense research for many years. While significant attention has been devoted to well-established allotropes such as fullerenes, carbon nanotubes, and graphene, the formation of carbon nanoparticles (CNPs) in dynamic environments such as flames and high-temperature fuel pyrolysis has recently emerged as a compelling area of cross-disciplinary investigation. CNPs exhibit unique optical and electrical characteristics, making them promising candidates for a diverse range of technological applications. These include, but are not limited to, enhanced performance electrodes in batteries, novel resistive switching devices (also in combination with metal oxide matrices), novel solar energy conversion, and more efficient fuel cells [1–4]. Moreover, CNPs generated within highly reactive hydrocarbon flame environments show direct evidence of quantum confinement effects [5], allowing their use as carbon quantum dots. Furthermore, the spectroscopic signatures of CNPs show striking similarities to spectral features observed in astronomical settings, suggesting a potential role as a fundamental constituent of cosmic dust, particularly within the interstellar medium (ISM), where they could influence radiative transfer and contribute to the formation of complex organic molecules [6,7,8].

Despite the considerable potential for technological exploitation and the intriguing connections to astrophysical phenomena, a comprehensive understanding of the fundamental process-structure-property relationships governing CNP formation remains a significant challenge. A common thread linking diverse phenomena, from the formation of nanocarbon materials in industrial processes to the presence of carbon dust in interstellar space, is the crucial role played in their formation/transformation by hydrogen atoms (H). The complex interplay between atomic hydrogen with graphene [9], as well as polycyclic aromatic hydrocarbons (PAHs), which serve as the molecular building blocks of CNPs [10], is hypothesized to be a key factor influencing both the pathways leading to specific carbon structures and their resulting optoelectronic and mechanical properties. Experimental evidence for the formation of covalent-like bonds both between aromatic units within individual PAH and between PAHs in the interlayer regions of CNP aggregates has been reported [11,12], although the precise nature and bonding mechanisms of these interlayer interactions remain a subject of ongoing investigation. These findings have led to the hypothesis that the formation of PAH interlayer CC bonds via $sp^3$ hybridization, leading to both "pancake" bonding (weak π-π stacking) and stronger covalent bonding, occurs through a molecular catalytic process [10] which can be facilitated by the presence of the H atoms that are readily available in reactive fluids such as flame environments. The presence of hydrogen can improve the structural and mechanical properties in analogy to what is observed in hydrogenated amorphous carbon films [13], where it promotes/stabilizes the $sp^3$-bonded carbon phase.

In previous works, exposure of carbon grains [14] and coronene [6] to H atoms was shown to lead to the formation of aliphatic CH bonds, as evidenced by the activation of CH stretching and bending bands in infrared spectra. The intensity of the CH bands increased with increasing flux of H atoms until saturation was reached, indicating that the hydrogenation process is limited by the number of available carbon sites. Here, to shed light on how the interaction with H atoms could promote carbon structural transformation, the effect of hydrogenation on flame-formed CNP films and single CNPs, generated under different conditions of flame equivalent ratio and residence time, was explored.

# 2. Experimental methods

Young and mature soot nanoparticles were produced from two premixed ethylene-air flames with a cold gas velocity of 9.8 cm/s and carbon-to-oxygen ratio (C/O) of 0.67 (equivalence ratio $\varphi$ = 2.01 – slightly rich flame) and 0.77 ($\varphi$ = 2.31 – rich flame) stabilized on a McKenna burner with a diameter of 6 cm.

Online and offline techniques using different sampling procedures were used for particle characterization (see Supplementary Material, SM, for more details. Preliminary particle size distribution (PSD) measurements were performed using a high dilution tubular sampling probe coupled to a Scanning Mobility Particle Sizer (SMPS) to guide the selection of the flame conditions (see SM for more detail).

The particles deposited on substrates were exposed to atomic H following several steps of exposure and subsequent IR measurements, up to the H atom fluence of $6.12 \times 10^{18}$ H atoms/cm$^2$. Atomic H was produced by dissociation of molecular hydrogen via microwave excitation (see reference [14] for more details), and hydrogenation was carried out in the same UHV chamber used for the IR spectroscopy measurements, maintaining a base pressure of $5 \times 10^{-8}$ mbar, without exposing the chamber to ambient pressure (more experimental details in SM). Raman spectroscopy, Atomic Force Microscopy (AFM), and macroscopic current vs. voltage (*IV*) measurements were performed prior to any hydrogenation and after the complete H exposure (more experimental details in SM). Scanning Tunneling Microscopy and Spectroscopy (STM/STS) were performed on isolated particles to investigate the local electronic properties of the soot particles, via both *IV* and d*I*/d*V* curves of the tunnelling current that flows between the tip and the sample surface acquired between −1.5 and 1.5 V. STM/STS was carried out in a RHK-VT scanning tunneling microscope (RHK Technology) at an UHV base pressure of $3 \times 10^{-11}$ mbar (similarly to [16], see

SM for more details). Atomic deuterium was produced by thermal dissociation of molecular deuterium, and hydrogenation of the sample was performed in the preparation chamber of the same UHV system (see [17] for more details). Over each particle investigated, ten curves were acquired and then averaged. Several steps of hydrogenation were performed, up to $1.27 \times 10^{15}$ D atoms/cm$^2$. The software Gwyddion [18] was used to process AFM and STM images.

## 3. Results and discussion

The PSDs presented in Figs. SM1 and SM2 clearly show that only particles smaller than 6 nm, classified as young soot, are present at HAB = 7 mm and 10 mm in the C/O = 0.67 flame. In contrast, particles larger than 20 nm, characteristic of mature soot, are observed at HAB = 10 mm and 14 mm in the 0.77 flame. Based on these findings, these flame conditions were selected for the subsequent investigation.

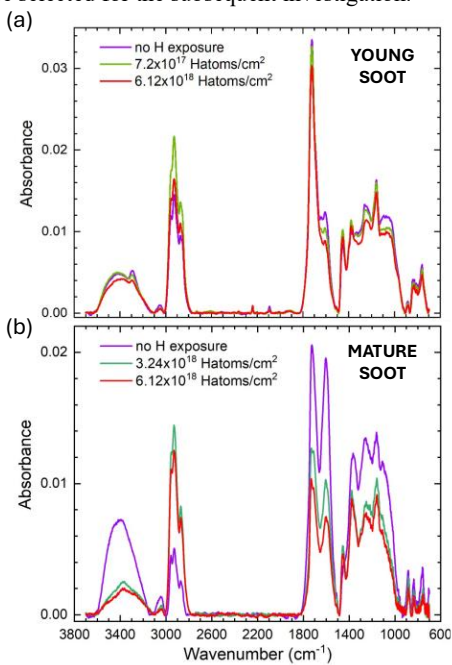

Fig. 1: Evolution of baseline-subtracted IR spectra of (a) young and (b) mature soot particles during hydrogenation at three representative steps of hydrogenation (all the spectra are shown in Fig. SM3).

First, the step-by-step evolution of hydrogen-induced chemical modification of CNP films was followed with IR spectroscopy, starting from the samples before hydrogenation and up to a total H fluence of $6.12 \times 10^{18}$ H atoms/cm$^2$. The baseline subtracted spectra, shown in Fig. 1, exhibit complex behavior: some bands show a monotonic trend, continuously decreasing or increasing with hydrogenation, whereas other bands display a trend inversion, firstly increasing and then decreasing with increasing exposure to atomic hydrogen. In the higher wavenumber range, from 3800 to 3100 cm$^{-1}$, a broad band centered around 3460 cm$^{-1}$ is observed, corresponding to the stretching vibrations of OH groups. A small peak, centered around 3280 cm$^{-1}$, is attributed to the stretching of alkyne sp CH bonds [19]. The subsequent region, from 3100 to 2800 cm$^{-1}$, is of great importance in our investigation, being dominated by the stretching of aromatic and aliphatic CH groups. The shape and intensity of these bands yield crucial information on the CH bonds in carbonaceous materials, making them essential for studying hydrogen adsorption and/or chemisorption mechanisms [11,13].

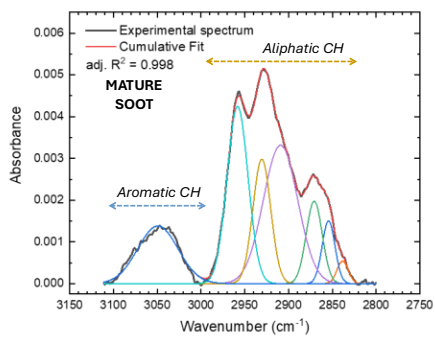

Fig. 2: Exemplary deconvolution of an IR spectrum recorded on mature soot particle film (before H exposure) in the aromatic/aliphatic CH stretches region (from left to right, the bands are assigned to aromatic CH and to aliphatic asym. CH$_3$, asym. CH$_2$, CH, sym. CH$_3$, sym. CH$_2$, and sym. CH$_2$ wing, as shown in Fig. SM4).

A detailed analysis of aromatic and aliphatic hydrogen was performed by deconvoluting the IR spectra in the CH stretching region (Fig. 2). The CH groups connected to aromatic structures are characterized by the band centered around 3045 cm$^{-1}$ [20], whereas the more complex region at lower wavenumber, due to CH aliphatic groups, was deconvoluted with multiple bands. In particular, the bands at 2960 and 2929 cm$^{-1}$ can be assigned to asymmetric CH$_3$ and CH$_2$ stretching, respectively, the band centered at 2903 cm$^{-1}$ to symmetric CH stretching, and the bands at 2871 and 2854 cm$^{-1}$ to symmetric CH$_3$ and CH$_2$ stretching, respectively. Finally, the small band at 2831 cm$^{-1}$ can be attributed to symmetric CH$_2$ stretching as well (usually called "wing" band) arising from aliphatic groups belonging to 5- and 6-member rings composed of at least two CH$_2$ groups [20,21]. The evolution of the integrated area of the aliphatic CH absorbance (which is proportional to the number of CH bonds) with increasing exposure to atomic H is shown in Fig. 3. We can observe that the two kinds of CNP films follow a similar initial increase of the aliphatic CH absorbance under increasing H flux, indicating that the H atoms are binding with the carbon atoms of the films, forming new CH, CH$_2$, and CH$_3$ bonds. The further exposure to atomic hydrogen produced a different effect: although for both young and mature soot samples the total area of the aliphatic CH reaches a maximum and then starts to decrease, young soot particles exhibit a faster and more critical reduction, possibly hinting at a higher reactivity towards interaction with H atoms and chemical and/or structural modifications. Indeed, the maximum value of the aliphatic CH area was reached after exposure to a fluence of around $0.8 \times 10^{18}$ H atoms/cm$^2$ for the young soot, while in the case of the mature particles the aliphatic CH area continues increasing up to the H fluence of about $3.2 \times 10^{18}$ H atoms/cm$^2$. The presence of a maximum in the aliphatic CH trend is different from what was seen in previous hydrogenation works on carbon grains [14], where the intensity of the CH bands increased with increasing flux of H atoms until saturation.

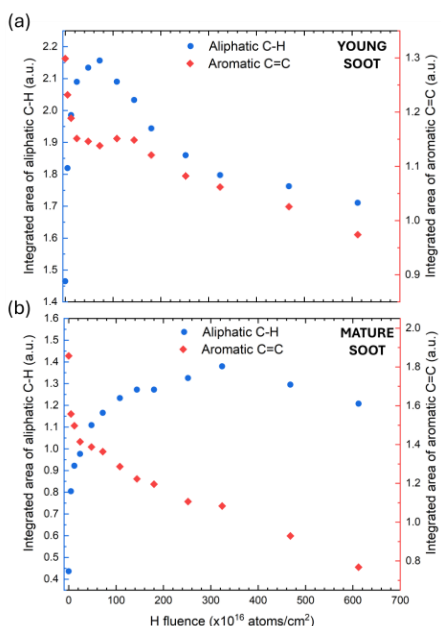

Fig. 3: Integrated area of total aliphatic CH bonds (from 3000 to 2810 cm$^{-1}$), blue circles, and aromatic C=C bonds (centered at 1600 cm$^{-1}$), red diamonds, for (a) young and (b) mature soot particles versus H fluence.

Moving to the lower wavenumbers, the two intense bands at around 1715 and 1600 cm$^{-1}$ are assigned to the stretching of the carbonyl group C=O and the stretching of sp$^2$ aromatic C=C bonds, respectively (as shown in Fig. SM5) [19]. The change of the intensity of the aromatic C=C band, evaluated as an integrated area of the 1600 cm$^{-1}$ band, upon exposure to H atoms, is shown in Fig. 3. Both samples show a similar general behavior, with a steep initial decrease, followed by a slower reduction. This reduction of the aromatic carbon can be associated with the increase of aliphatic CH bonds: atomic H has been shown to break the sp$^2$ C=C bonds and form CH$_n$ bonds [11,14]. Notably, for the young soot, a plateau is observed at $0.8$-$1.2 \times 10^{18}$ H atoms/cm$^2$, coinciding with the H fluence where the aliphatic CH intensity begins to drop. In Fig. 3, a similar but less distinct plateau is seen in the reduction of the aromatic C=C area for the mature soot at around $3.0 \times 10^{18}$ H atoms/cm$^2$.

The region between 1500 and 1000 cm$^{-1}$, often called the fingerprint region, is characterized by several complex overlapping bands arising from various molecular vibrations, including CO group stretches (such as esters), CH and OH groups bending of various types, and CC single bond stretches. Due to this complexity and band overlap it is less reliable for identifying functional groups. As shown in Fig. 1, the general intensity of this region decreased with increasing H exposure, but a more precise and deeper deconvolution analysis is needed to separately follow individual bands. In a preliminary analysis, after the subtracting of a local multipoint linear baseline, the peaks centered at 1450 and 1380 cm$^{-1}$, assigned to the mixed aliphatic CH$_2$ and CH$_3$, and the CH$_3$ alone bending, respectively, exhibited a similar trend to their corresponding stretching at around 2900 cm$^{-1}$, showing an initial increase followed by a reduction in intensity. In contrast, the peaks at around 1250 and 1130 cm$^{-1}$, assigned to the bending of CO and aromatic CH groups, showed a continuous decrease.

In summary, IR spectroscopy revealed an initial increase in the aliphatic CH bonds followed by a decrease, mostly prominent in the younger soot sample. This transient behavior suggests an initial hydrogenation of the available carbon sites, followed by a rearrangement of the hydrogenated structures. Concurrently, the continuous decrease of the aromatic C=C signal in the IR spectra indicates a progressive disruption of the aromatic sp$^2$ network. This excludes the possibility of graphitization, i.e., the growth in size of aromatic domains, which would instead display a stabilization or even an enhancement of the C=C band. Moreover, no evidence of an increase in other chemical bonds was found. Finally, given the vacuum conditions maintained throughout the experiment ($10^{-4}$-$10^{-8}$ mbar), oxygen incorporation was negligible, excluding oxidation as a contributing factor. Therefore, we propose that the observed effects are mainly due to hydrogen-induced cross-linking. However, their vibrations would fall in the complex fingerprint region [22] due to the non-polar and likely symmetric nature of sp$^3$ CC single bonds, they are not directly observable via IR spectroscopy. To overcome this limitation, other characterization techniques were utilized to explore the chemical, structural, mechanical, and electrical modifications induced by hydrogenation with atomic H.

Figure 4 shows the Raman spectra of young and mature soot particle films after subtraction of the fluorescence background and normalization to the maximum of the G band at about 1600 cm$^{-1}$, before and after the hydrogenation. Both samples exhibited the characteristic D and G bands associated with disorder and sp$^2$-hybridized carbon, respectively [23]. Table 1 summarizes some key parameters derived from the Raman spectra measured before and after hydrogenation: the G peak position, the $I$(D)/$I$(G) ratio, $m$/$I$(G) (where $m$ is the slope of the fluorescence linear background identified in the Raman spectra).

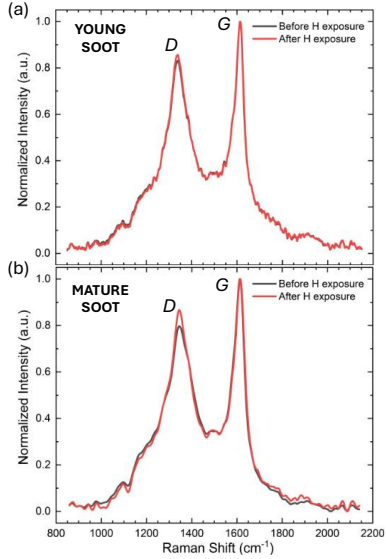

Fig. 4: Normalized Raman spectra of CNPs before and after H exposure, acquired from: (a) young and (b) mature soot particle films, before (black lines) and after (red) H exposure.

We can observe that hydrogenation resulted in a decrease in fluorescence, probably because of the breaking of the aromatic structures [24], evidenced by the aromatic C=C decrease in the IR spectra. Moreover, while the G band position remains fixed, the $I$(D)/$I$(G) ratio increased, indicating a variation in the structural configuration. In the literature, an increase of $I$(D)/$I$(G) has been associated with either amorphization (e.g., fragmentation of the aromatic network in graphite and graphene) or increased short-range ordering in amorphous carbon (e.g., a higher probability of finding 6-fold rings), depending on the initial structure and on the transformation trajectory [23]. However, both amorphization and ordering are associated with a shift in the position of the G peak [25], which arises from in-plane stretching vibrations. Here, instead, the G position remains almost unchanged, suggesting a structural reconfiguration, which mostly affects the interlayer structure, such as an out-of-plane CC cross-linking bond. This interpretation is supported by reports of a similar modest increase in $I$(D)/$I$(G) without G band shift, attributed to the formation of $sp^3$ carbon cross-links between carbon nanotubes [26].

To support our interpretation of the hydrogen-induced interlayer structural modifications, the nano-mechanical properties of the CNP films were investigated. Nanoindentation force vs. distance curves were acquired on the CNP films before and after the hydrogenation process (see SM for the film morphology imaged with AFM – Fig. SM6 – and more details on the evaluation of nano-mechanical properties). From the fit of the nanoindentation curves (shown in Fig. SM7) we obtained the mean Young's modulus, $E$, reported in Tab. 1. These values are in good agreement with the results reported in the literature obtained in similar samples [15], with a higher Young's modulus for more mature and graphitic CNPs. After hydrogenation, the $E$ increased for both samples. In literature, an increase in $E$ (i.e., the rigidity or stiffness of a material) was shown to correlate with a higher level of cross-linking [27]. Moreover, a higher Young's modulus was reported to correlate with a decrease in the lattice fringe spacing of diesel soot particles [28], that would be expected in case of increased cross-linking between the layers.

To further investigate the possible interlayer or interparticle distance reduction after hydrogen-induced cross-linking, the macroscopical electrical properties of the CNP films were studied by acquiring $IV$ characteristics curves (shown in Fig. 5) on the thin film produced by deposition of mature particles on an IDE gold substrate. As H exposure increased, the maximum current flowing through the sample increased. while the shape of the curves remained non-linear. Previous studies have suggested that the electrical conduction in CNP films can be ascribed to tunneling and percolation effects [29]. The observed increase in maximum current could be attributed to either improved tunneling between the conductive regions in the film, resulting in improved percolation paths, or to an increased level of graphitization. However, based on the IR spectroscopy results, graphitization can be ruled out. In multilayer carbon nanomaterials tunnelling current has been shown to strongly depend on the interlayer distance, increasing as the distance decreases [30]. Therefore, the higher current likely indicates a lower interlayer distance (e.g., from 3.4 Å to about 1.4 Å [31]) or interparticle distance, particularly near newly formed cross-linking bonds.

Table 1: Comparison of the relevant parameters extracted from Raman spectroscopy and nano-mechanical analyses before and after hydrogenation (average ± SD).

|  | Young soot particles | | Mature soot particles | |
| --- | --- | --- | --- | --- |
|  | Before H | After H | Before H | After H |
| $m$/$I$(G) (μm) | 4.1 ± 0.3 | 3.9 ± 0.5 | 3.2 ± 0.3 | 2.5 ± 0.1 |
| G position (cm$^{-1}$) | 1615 ± 1 | 1615 ± 2 | 1615 ± 1 | 1614 ± 2 |
| $I$(D)/$I$(G) | 0.82 ± 0.02 | 0.85 ± 0.02 | 0.80 ± 0.02 | 0.86 ± 0.02 |
| $E$ (GPa) | 4.5 ± 0.4 | 6.3 ± 0.4 | 5.0 ± 0.3 | 6.6 ± 0.2 |

Before hydrogenation, young soot particles on substrates with flat gold terraces were first identified in STM mode (Fig. SM8), and then, *IV* curves were recorded via STS by positioning the tip approximately over the center of the particle, at a tunnelling distance. Fig. SM9 shows exemplary *IV* curves measured. As shown in Fig. 6(a), for young soot, the distribution of the maximum current (at 1.5 V) can be described with a normal distribution curve, with a $<I_{MAX}>$ of 0.24 nA and a standard deviation SD = 0.23 nA. After exposure to atomic D, CNPs were first visualized in STM, and *IV* curves were then acquired, obtaining a mean value of $<I_{MAX}>$ = 1.7 nA and a SD = 0.7 nA. Simultaneously, the differential conductance (d*I*/d*V*) curves were acquired by a lock-in amplifier as a function of bias voltage. The resulting spectra produce information about the local electronic density of states (LDOS) and, therefore, the electronic band gap ($E_{GAP}$). As shown in Fig. 6(b), before hydrogenation, the $E_{GAP}$ values largely spread between 0 and 2.2 eV (normal $<E_{GAP}>$ = 1.0 eV and SD = 0.6 eV). After D exposure, smaller $E_{GAP}$ values were measured (normal $<E_{GAP}>$ = 0.09 eV, SD = 0.20 eV), with 80% of particles showing $E_{GAP}$ = 0 eV (semi-metallic behavior – see Fig. SM10). In literature, CNPs have been shown to exhibit quantum confinement, with the band gap decreasing with increasing dimension [5]. However, in this study, all the studied particles have similar dimensions (all smaller than 6 nm, in agreement with the PSDs), suggesting that the observed reduction in band gap is not size-related, but rather due to a hydrogen-induced chemical modification. This transformation from a semiconductive to a semi-metallic nature can be explained by considering hydrogen atoms to bond to the carbon network, inducing chemical and structural modifications and creating new accessible electronic states within the original valence-conduction bandgap of the carbon nanoparticles (effectively closing the gap) [32].

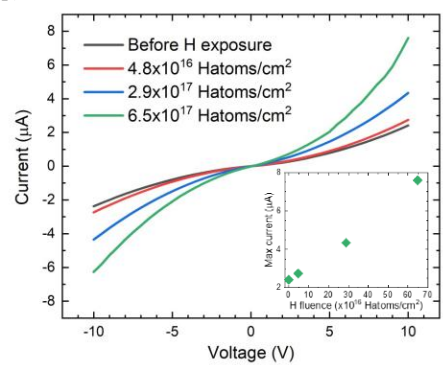

Fig. 5: Evolution of current vs. voltage (*IV*) curves recorded on the mature soot particle film before and after subsequent steps of hydrogenation. The inset shows the maximum current values measured at 10 V.

We propose that the observed changes in mechanical and electronic properties of CNPs are primarily driven by hydrogen-induced cross-linking. A likely mechanism involves atomic H generating radical sites on polycyclic aromatic hydrocarbon units, facilitating stacking and clustering through interactions involving unpaired electrons delocalized across open-shell structures. While previous studies have described the dynamic character of super-hydrogenation in PAHs, where incoming H atoms can both add and abstract, forming a $H_2$ molecule [11], here we demonstrate its structural and electronic consequences in flame-formed nanoparticles. The creation of open-shell intermediates enables CC single-bond formation and progressive cross-linking, evidenced by stiffening of the material and an enhancement of electrical conductivity. This interpretation is consistent with previous reports of hydrogen-induced covalent coupling upon collision of super-hydrogenated PAHs at contiguous radical-like sites, which produces dendritic nanostructures [33]. However, the present work extends these findings by experimentally linking nanoscale chemical processes to macroscopic mechanical and electronic properties.

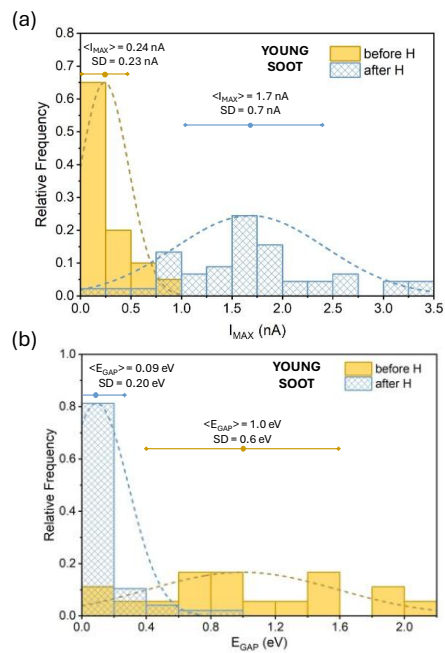

Fig. 6: Distribution, as relative frequency, of (a) the maximum current measured at 1.5 V from the IV curves and (b) the EGAP measured from the dI/dV curves on CNPs before (yellow filled bins) and after (patterned blue bins) hydrogenation. Normal distribution curves are shown as dashed lines, and the mean and SD values are indicated.

A further key finding is the non-monotonic evolution of aliphatic CH bonds, observed in IR spectra. In sharp contrast to the monotonic saturation trends reported in earlier studies, we observe an initial increase followed by a sharp decline in aliphatic CH intensity, particularly in young particles, which were previously shown to possess a higher radical density [34]. In these samples, hydrogenation led to a significant chemical transformation and band gap closure, as revealed by STS measurements. This provides the first experimental evidence of a hydrogen-induced transition from semiconducting to semi-metallic behavior in individual soot nanoparticles. Moreover, it suggests that the reactivity and initial structure of the soot play a crucial role in the hydrogenation process. Mature soot, which displays reduced reactivity, undergoes limited chemical or electronic change, responding more like previously studied graphitic materials.

## 4. Conclusions

This study investigated the effects of hydrogenation on CNP films, combining spectroscopic, mechanical, and electrical characterization techniques. Our findings reveal a complex interplay of structural and electronic modifications induced by hydrogenation.

Raman spectroscopy and AFM revealed structural modifications upon hydrogen exposure, while nanoindentation measurements indicated a significant increase in Young's modulus, consistent with enhanced interlayer cross-linking. Macroscopic and local electrical measurements ($IV$ and STS curves) revealed a significant increase in conductivity, accompanied by a substantial reduction in the electronic band gap, with most particles exhibiting semiconducting behavior after hydrogenation. These results, supported by IR spectroscopy and STS measurements, confirm that the observed changes are not due to increased graphitization or size effects, but rather to H-induced chemical bonding that introduces new electronic states and reduces interparticle and interlayer distances. We propose hydrogen-induced cross-linking as the central mechanism responsible for these changes: atomic hydrogen generates radical sites on PAHs, promoting stacking and clustering via delocalized unpaired electrons in open-shell structures.

These findings expand upon earlier work by offering new insights into the consequences of hydrogenation in combustion and astrophysical environments, while at the same time suggesting pathways for the tailored design of carbon-based nanomaterials with enhanced mechanical and electronic properties. Moreover, future studies on the stability and/or reversibility of hydrogen-induced structural and chemical modifications of flame-formed CNPs under varying ambient conditions (e.g., air exposure, temperature, pressure) would further extend the insights gained in this work.

### Novelty and Significance Statement

The novelty of this research lies in the comprehensive investigation of the impact of atomic hydrogen exposure on flame-formed carbon nanoparticles (CNPs), specifically detailing the complex, non-monotonic evolution of aliphatic CH bonds and the concurrent decrease in aromatic C=C bonds. This study uniquely correlates these chemical changes with significant increases in mechanical stiffness (Young's modulus) and macroscopical and local electrical conductivity, providing multi-technique evidence for a hydrogen-induced CC cross-linking mechanism. These findings are crucial because they offer valuable insights into the fundamental process-structure-property relationships governing CNP evolution in hydrogen-rich environments. This understanding has direct implications in soot formation in combustion and carbon grains in interstellar media, as well as for the tailored design of carbon-based materials with enhanced mechanical and electrical properties for various technological applications.

**Author Contributions**

**Luca Basta**: Formal analysis, Investigation, Writing - Original Draft. **Francesca Picca**: Formal analysis, Investigation, Writing - Original Draft. **Pegah Darvehi**: Formal analysis, Investigation, Writing - Original Draft. **Vincenzo Pagliara**: Formal analysis, Investigation, Writing - Original Draft. **Alberto Aloisio**: Supervision, Founding Acquisition, Writing - Original Draft. **Mario Commodo**: Conceptualization, Supervision, Writing - Original Draft. **Patrizia Minutolo**: Supervision, Writing - Original Draft. **Vito Mennella**: Conceptualization, Supervision, Writing - Original Draft. **Stefan Heun**: Conceptualization, Supervision, Writing - Original Draft. **Stefano Veronesi**: Conceptualization, Formal analysis, Investigation, Supervision, Writing - Original Draft. **Andrea D'Anna**: Conceptualization, Founding Acquisition, Supervision, Writing - Original Draft.

**Declaration of competing interest**

The authors declare that they have no known competing financial interests or personal relationships that could have appeared to influence the work reported in this paper.

**Acknowledgements**

The authors would like to thank R. Guastafierro for support in the IR experiments. L. Basta and A. Aloisio acknowledge funding by the NextGenerationEU European initiative through the Italian Ministry of University and Research, PNRR, project "Centro Nazionale HPC, Big Data e Quantum Computing Italian Center for Super Computing (ICSC)" – Spoke 1, MUR: CN_00000013, CUP UNINA:

E63C22000980007. This material is based upon work supported by the Air Force Office of Scientific Research under award number FA8655-25-1-7003.

**Supplementary material**

Supplementary material is available, showing complementary experimental methods details and data.

# Supplementary Material

# Impact of hydrogenation on the structure, chemistry, and electrical properties of flame-synthesized carbon nanoparticle films

Luca Basta[a,b,*], Francesca Picca[b], Pegah Darvehi[c], Vincenzo Pagliara[d], Alberto Aloisio[a], Mario Commodo[c,*], Patrizia Minutolo[c], Vito Mennella[d], Stefan Heun[e], Stefano Veronesi[e], and Andrea D'Anna[b]

[a]Dipartimento di Fisica Ettore Pancini, Università degli Studi di Napoli Federico II, Complesso Universitario di Monte S. Angelo, Edificio 6, Via Cinthia, 21, Napoli 80126, Italy
[b]Dipartimento di Ingegneria Chimica, dei Materiali e della Produzione Industriale, Università degli Studi di Napoli Federico II, P.le Tecchio 80, Napoli 80125, Italy
[c]Istituto di Scienze e Tecnologie per l'Energia e la Mobilità Sostenibili, Consiglio Nazionale delle Ricerche, P.le Tecchio 80, Napoli 80125, Italy
[d]Istituto Nazionale di Astrofisica - Osservatorio Astronomico di Capodimonte, Via Moiariello 16, Napoli 80131, Italy
[e]NEST, Istituto Nanoscienze - CNR, and Scuola Normale Superiore, Piazza San Silvestro 12, 56127 Pisa, Italy

___

## *On-line particle characterization*

*Sampling methods*

For online characterization techniques, the particles produced in the flames were collected by a tubular steel sampling probe with an orifice located downward and at the centerline of the flame. The particles were diluted immediately after entering the orifice, using nitrogen gas with a flow rate of 30 L/min. A probe with an orifice diameter of 0.2 mm was used, achieving a dilution ratio of approximately 1:3400 to strongly reduce particle coagulation. The height above the burner (HAB) where the probe orifice is positioned defines the particle residence time in the flame. To account for perturbations introduced by the probe, the effective HAB is considered to be 3-4 mm lower than the actual probe position [1].

For this study, soot samples were classified as young or mature, without considering variations between sampling HABs in each flame, which were only minimal (as shown in the PSDs below). For both flames, a slightly higher sampling position was chosen when a higher soot volume fraction was required in order to shorten the sampling time or to minimize mechanical and thermal stress on the delicate gold contacts (50 nm thick, 10 μm wide) of the IDE substrates.

*Particle size distributions (PSDs)*

The particle size distribution (PSD) functions were measured using a Scanning Mobility Particle Sizer SMPS (TSI Incorporated), consisting of an X-ray charger (Model 3088), an electrostatic classifier (Model 3082), and an ultrafine condensation particle counter (Model 3776). Initially, the Multi-Instrument Manager software was used to correct for diffusion losses and the presence of multiple charged particles within the instrument's internal pathways. PSDs were further adjusted both to account for the dilution ratio and for particle losses along the sampling line.

Volume size distributions of the particles collected from the flame at a carbon-to-oxygen ratio C/O = 0.67 and heights above the burner HABs of 7 mm and 10 mm are shown in Fig. SM1. Under these conditions, only particles smaller than 6 nm, which can be referred to as young soot particles, are measured. In contrast, particles collected from the flame with C/O = 0.77 at HABs = 14 mm and 18 mm are larger than 10 nm, indicating the presence of mature soot particles.

The volume PSDs of the carbon nanoparticles were fitted with lognormal distribution functions:

$$\frac{dV}{dLogD_m} = \frac{A}{\sqrt{2\pi}wD_m} e^{\frac{-\left(\ln{D_m}/{VMD}\right)^2}{2w^2}}$$

where $w$ is the log standard deviation, and $VMD$ is the volume median diameter [3]. For CNPs generated in the C/O = 0.67 flame, a bimodal volume size distribution was observed at HAB = 7 mm and 10 mm. Accordingly, two lognormal functions were needed to fit the data. The first mode exhibited a consistent volume mean diameter ($VMD$) of about 2.0 nm at both HABs. In contrast, the second mode extended slightly towards larger diameters and became broader with increasing HAB. Nevertheless, all detected particles remained smaller than 6 nm.

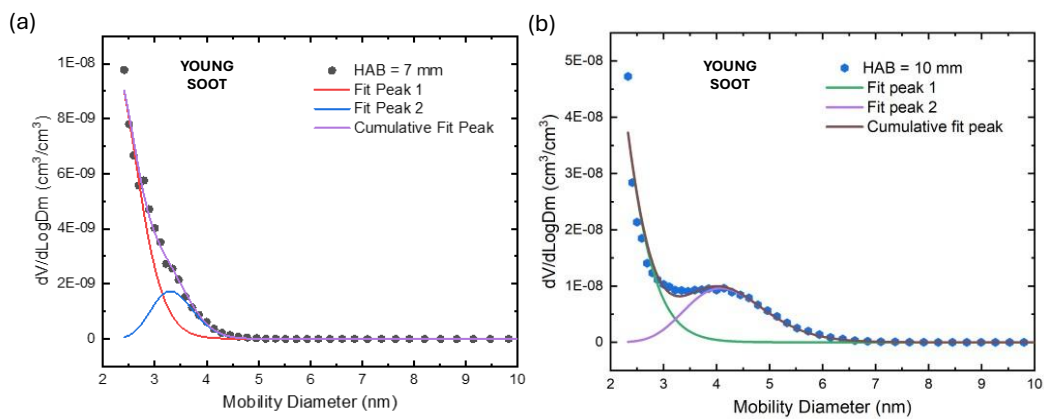

**Figure SM1**. Volume PSDs with the fit curve for young soot particles sampled in the flame with C/O=0.67 at (a) HAB = 7 mm and (b) HAB = 10 mm. The lognormal fitting curves, together with the cumulative fits, are shown.

As shown in Fig. SM2, CNPs from the richer flame (C/O = 0.77) exhibited a similar unimodal size distribution at both sampling heights. A single lognormal function was sufficient to fit each PSD. At both HABs, the volume size distributions consisted of particles larger than 20 nm, with $VMD$ of the order of 100 nm (see Fig. SM2).

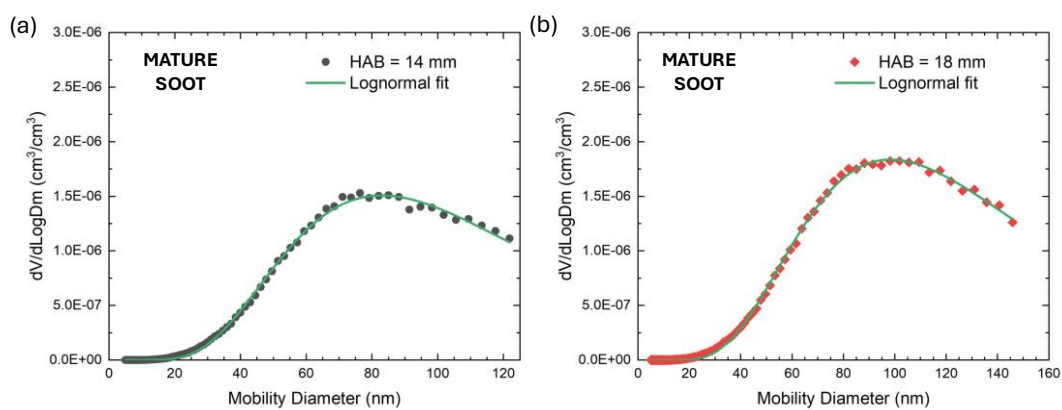

**Figure SM2**. Volume PSD with the fit curve (green line) for mature soot particles collected in the flame with C/O = 0.77: (a) at HAB = 14 mm and (b) at HAB = 18 mm.

When comparing particle characterization results obtained using different techniques that rely on distinct sampling methods, it is important to account for perturbations introduced by the sampling system itself. Specifically, the flame region probed using a tubular probe was shifted by about 4 mm relative to the region analyzed via thermophoretic sampling. This shift is attributed to changes in temperature and flow caused by the presence of the probe in the flame [1]. As a result, the young soot particles collected in the flame with C/O = 0.67 at HAB = 7 mm using probe sampling correspond to those collected by thermophoretic sampling at HAB = 3 mm. Likewise, mature soot collected with the tubular probe at HAB = 14 mm and HAB = 18 mm corresponds to the mature soot collected by thermophoretic sampling at HAB = 10 mm and 14 mm.

## Off-line particle characterization

### Sampling methods

For offline analysis, particles were deposited onto three different substrates: atomically flat gold on mica (prepared as described in [2]), cesium iodide (CsI), and InterDigitated-Electrode (IDE) substrates using two deposition methods: thermophoretic and impact sampling. In thermophoretic sampling, the substrate was inserted one or multiple times into the flame for 30 ms each, through a pneumatic actuator to capture particles exploiting thermal forces. Thermophoretic sampling was used in the 0.77 flame to produce mature soot particle films on the CsI substrate (230 insertions at HAB = 10 mm) and on IDE (30 insertions at HAB = 14 mm). Furthermore, to collect isolated particles for Scanning Tunneling Microscopy and Spectroscopy (STM/STS) measurements, young soot particles were sampled onto gold on mica using a single insertion of the pneumatic actuator at HAB = 3 mm in the 0.67 flame. In impact sampling, particles collected with the tubular probe (with an orifice diameter of 2.5 mm, resulting in a dilution factor of approximately 1:20) were deposited directly onto the substrate, which was placed inside a closed holder located upstream of the probe. This method was employed for the 0.67 flame to collect enough material, avoiding a number of insertions too high. Films of nanoparticles were thus obtained on the CsI substrate, with a collection time of 240 minutes at HAB = 10 mm.

*Infrared (IR) spectroscopy*

Transmission IR spectroscopy was carried out using a Bruker Vertex 80v FT spectrometer with a UHV chamber positioned between source and detector. IR access was provided by two optical windows, and the beam path was purged with nitrogen to eliminate air and water interference. Samples were mounted on a magnetically coupled translator, enabling vertical alignment and rotation between the two positions for H exposure and IR measurement without breaking the vacuum. Measurements were conducted over the spectral range of 4000-500 cm$^{-1}$ with a resolution of 2 cm$^{-1}$.

Fig. SM3 shows the full evolution of the absorbance IR spectra collected on CNP films with exposure to atomic hydrogen.

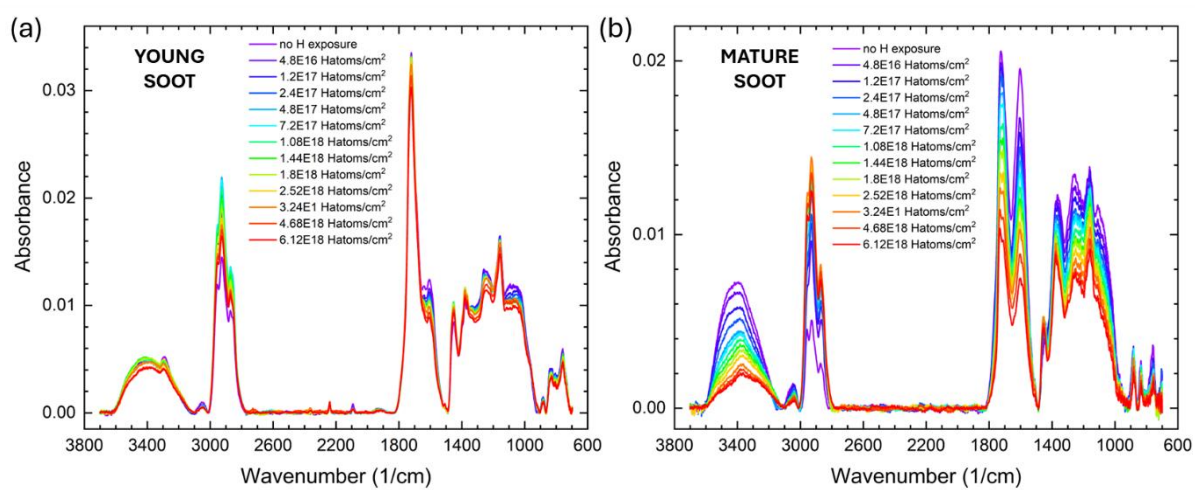

**Figure SM3**. Evolution of baseline-subtracted IR spectra of (a) young and (b) mature soot particle films during hydrogenation.

Fig. SM4 shows the deconvolution of an exemplary IR spectrum collected in the region between 3100 cm$^{-1}$ and 2800 cm$^{-1}$, where the bands arising from the CH$_n$ groups (stretching) bonded to aromatic and aliphatic structures are visible.

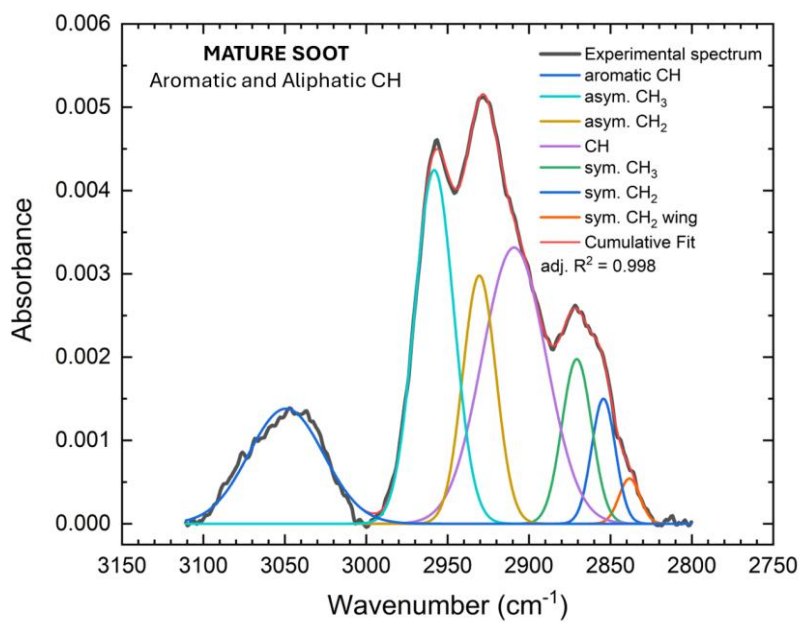

**Figure SM4**. Exemplary deconvolution of the IR spectrum recorded on a mature soot particle film (before H exposure) in the aromatic/aliphatic CH stretches region. The assignment of each band is shown.

Fig. SM5 shows the IR spectral region where the C=O and aromatic C=C stretching vibrations appear. As a first approximation, the spectrum can be deconvoluted into two bands corresponding to these two types of bonding, yielding good agreement with the experimental data (adj. $R^2$ = 0.994).

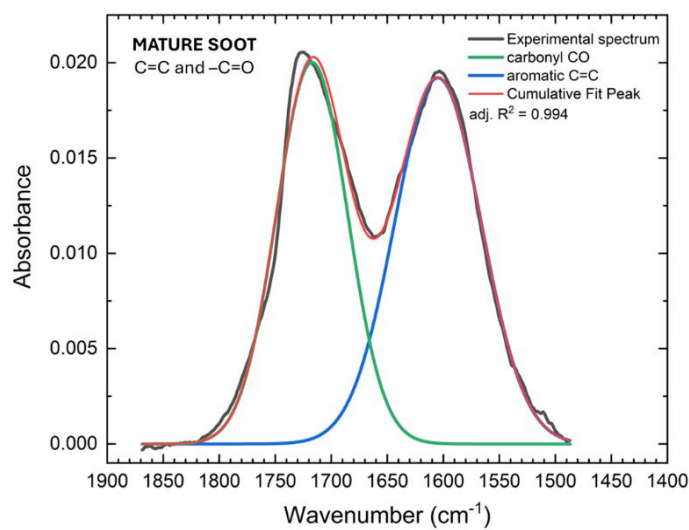

**Figure SM5**. Exemplary deconvolution of the IR spectrum recorded on a mature soot particle film (before H exposure) in the carbonyl C=O and aromatic C=C stretches regions.

*Raman spectroscopy experimental details*

Raman spectra were acquired using a Horiba XploRA Raman microscope system equipped with a 100× objective (NA 0.9, Olympus) and a frequency-doubled Nd:YAG laser ($\lambda$ = 532 nm).

To prevent structural changes in the sample due to thermal decomposition and ensure optimal resolution, laser beam power, exposure time, and other instrumental parameters were carefully selected. Spectra were acquired with a laser beam power set at 10% attenuation, and an accumulation (exposure) time of seven cycles, each lasting 40 seconds. Ten randomly selected spots per sample were averaged to enhance statistical relevance, and all spectra were subsequently baseline-corrected and normalized to the maximum of the G peak at approximately 1600 cm$^{-1}$.

*Atomic Force Microscopy (AFM) and nanoindentation force vs. distance curves*

The Scanning Probe Microscope NTEGRA Prima (NT-MDT) was used for imaging in semi-contact mode and for nanoindentation, following a similar procedure as in [6].
A diamond-coated tip (DT-NCHR from NT-MDT) was used for this study, with a tip radius of around 150 nm, a nominal spring force constant of 150 N/m, and a resonant frequency of 371 kHz. Force vs. distance curve measurements were performed, evaluating the interaction between the sample and the probing tip. Calibration with a freshly cleaved HOPG sample from NT-MDT was performed to extract the cantilever and tip parameters required for the analysis. Several curves were acquired (at least 10 for each sample) and then averaged.

Fig. SM6 shows the typical granular morphology of soot particle films made by thermophoretic deposition of young soot particles, acquired with AFM in semi-contact mode.

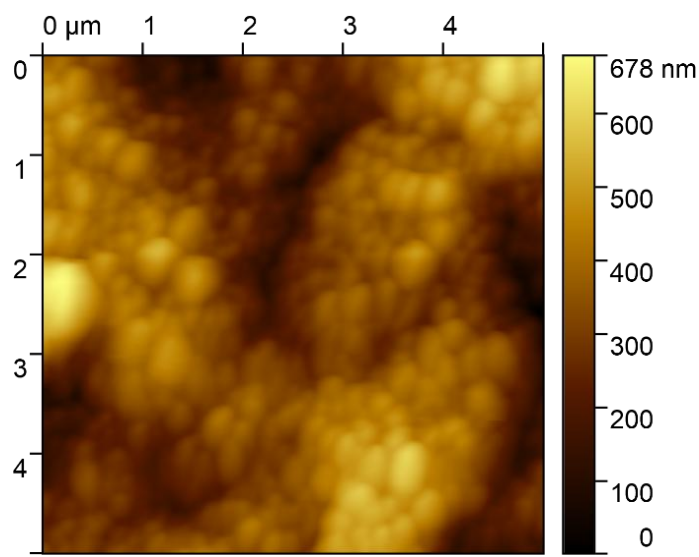

**Figure SM6**. Exemplary AFM semi-contact height image collected over an area of 5 × 5 μm$^2$ on the films of young soot particles.

As shown in a previous work [4], AFM can be operated in force spectroscopy mode to extract the nano-mechanical properties of the surface of soot films. When the cantilever spring constant and the photodetector sensitivity are known, the interactive force between the AFM tip and the sample can be recorded while approaching and retracting the tip. Initially, at high tip-sample distance, the interactive force is zero. As the tip approaches the sample, a "jump-to-contact" instability arises, causing a rapid cantilever deflection once the force gradient surpasses the spring constant. After the sample approach, retraction of the probe from the sample surface results in sustained contact due to adhesive forces. The "pull-off" instability in the curve is observed upon the spring constant exceeding the force gradient at the point where $F$ exhibits the negative minima corresponding to $F_{adh}$. The application of continuum elastic theory allows for the determination of the sample's mechanical properties. The Hertz theory assumes a spherical or paraboloidal indenter acting on a semi-infinite plane under conditions of negligible adhesion. Based on Hertz theory, the Derjaguin–Müller–Toporov (DMT) model represents a refinement of the Hertzian approach by incorporating the contribution of adhesive forces [5].

The following equation:
$$F = 4/3 \cdot E^* \cdot \sqrt{Rd^3} + F_{adh}$$
where $R$ is the tip radius, $d$ the sample-tip distance, and $F_{adh}$ the maximum adhesion force (the negative minima at $d = 0$), allows to obtain the effective elastic modules, defined as in [6]. The diamond-coated probes utilized in this study were characterized by a Young's modulus $E_{tip} =$ 1140 GPa and a Poisson's ratio $\upsilon_{tip} = 0.07$ [7]. Determination of the sample's Young's modulus $E_{film}$ was achieved by fitting the low loading force stages of the retraction curve, which corresponds to the elastic regime, and using the same Poisson ratio as for HOPG, $\upsilon_{HOPG} = \upsilon_{film} = 0.3$ [6].

Fig. SM7 shows force vs. distance retraction curves recorded on the CNP films.

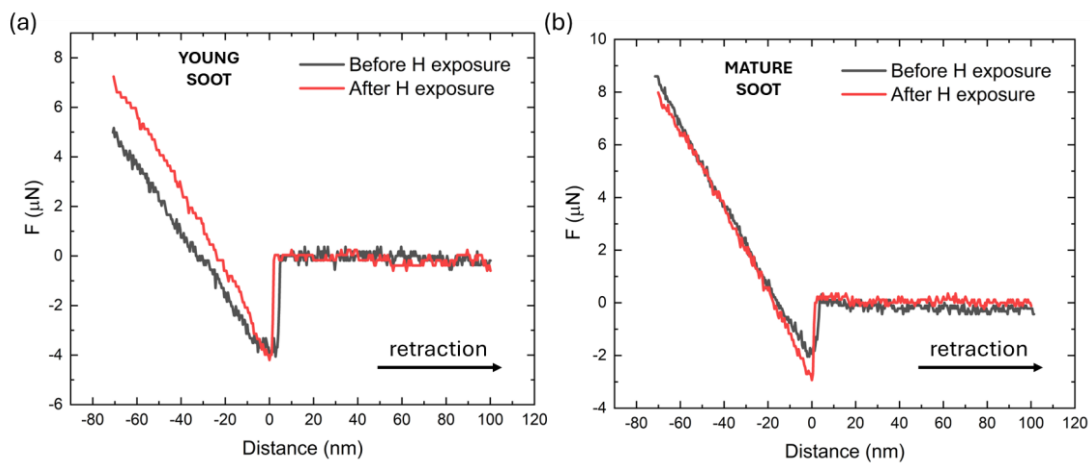

**Fig. SM7**: Force vs. distance retraction curves recorded before (black lines) and after (red lines) hydrogenation on (a) young and (b) mature soot particle film samples.

*Scanning Tunneling Microscopy and Spectroscopy (STM/STS)*

The STM tips are fabricated by electrochemically etching a tungsten wire, following a previously established procedure detailed in [8]. This optimized method produces tips with a diameter smaller than 20 nm [9], enabling a typical lateral resolution in STM of around 0.1 nm, given that the tunneling current is primarily localized near the point closest to the substrate. Before use, the prepared tips undergo overnight degassing followed by flashing to eliminate tungsten oxide. For STM imaging, the tip is grounded, and the reported bias values are applied to the sample. Images are acquired in constant current mode, where the tip's vertical movement is adjusted by a feedback loop to maintain a constant tunneling current.

Fig. SM8 shows the morphology of single soot particles on atomically flat gold terraces measured by STM imaging.

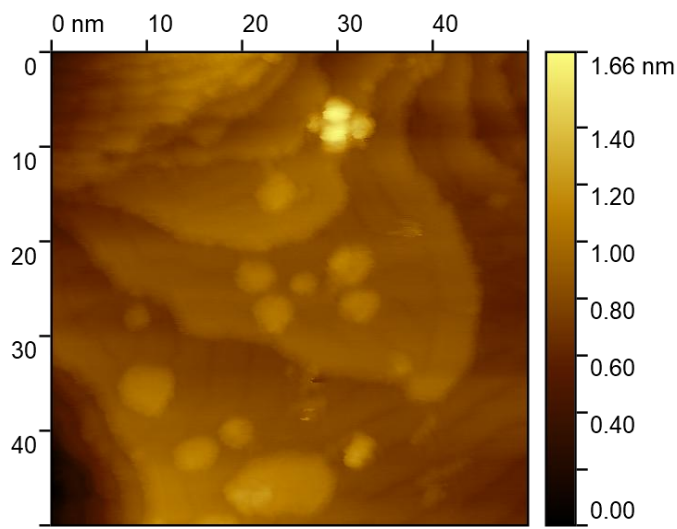

**Figure SM8**. Exemplary morphology imaged with constant current STM in an area of 50 × 50 nm², showing just-formed young soot particles over atomically flat gold terraces (setpoint $I = -0.70$ nA, bias voltage $BV = -1.3$ V).

Fig. SM9 shows two exemplary tunnelling current/voltage bias curves recorded with STS on single young soot particles before and after hydrogenation.

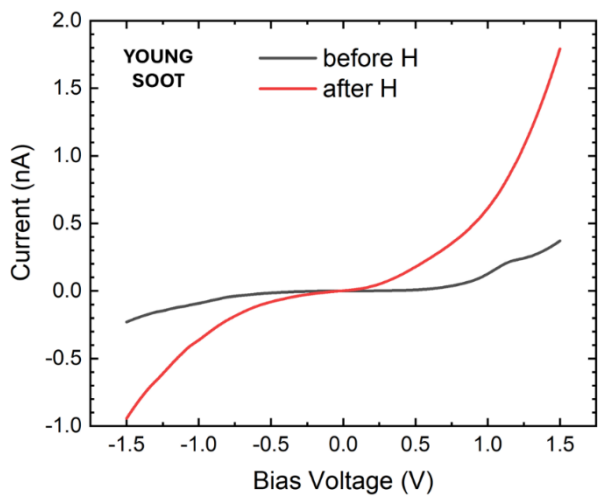

**Figure SM9**. Exemplary $IV$ curves acquired between $-1.5$ V and $1.5$ V with STS on isolated young CNPs before (black curve) and after (red curve) exposure to atomic hydrogen (deuterium).

Fig. SM10 shows exemplary differential conductance ($dI/dV$) spectra measured on CNPs before and after hydrogenation. The CNPs before H exposure show a semiconductive nature and an $E_{GAP}$ of 0.7 eV, while the CNPs after hydrogenation exhibit semi-metallic behavior, with no band gap.

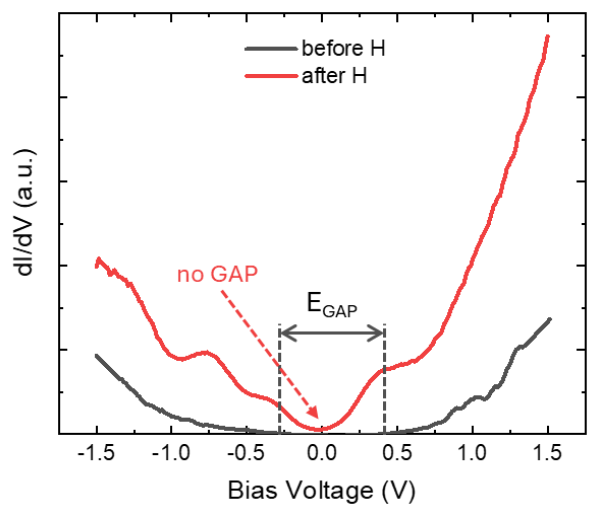

**Figure SM10**. Exemplary d*I*/d*V* curves acquired between −1.5 V and 1.5 V with STS on isolated young CNPs before (black curve) and after (red curve) exposure to atomic hydrogen (deuterium). The electronic band gap is indicated.

*Current vs. Voltage (*IV*) measurements*

For *IV* measurements, particles were deposited onto InterDigitated-Electrode (IDE) substrates (Micrux ED-IDE1-AU) using thermophoretic sampling. To deposit sufficient material and achieve a uniform film, the substrate was inserted into the 0.77 flame 30 times for 30 ms each using a pneumatic actuator. After deposition, the IDE substrate was mounted on a Linkam HFS600E-PB4 stage, and *IV* curves were recorded using a Source Measure Unit (model 200, Ossila Ltd., Sheffield, UK). Voltage sweeps were performed in the range −10 and 10 V, at a scan rate of 6.2 V/s.